\documentclass[english,12pt,preprint]{elsarticle}
\usepackage[T1]{fontenc}
\usepackage[latin9]{inputenc}
\usepackage{babel}

\usepackage{verbatim}
\usepackage{amsmath}
\usepackage{graphicx}
\usepackage{amssymb}

\journal{Physics Letters B}
\usepackage{bbm}
\usepackage{bm}
\newcommand{\sla}[1]{\rlap{$#1$}/}

\begin{document}
\global\long\def\abs#1{\left| #1 \right| }
\global\long\def\half{\frac{1}{2}}
\global\long\def\partder#1#2{\frac{\partial#1 }{\partial#2 }}
 \global\long\def\comm#1#2{\left[ #1 ,#2 \right] }

\global\long\def\Tr#1{\textrm{Tr}\left\{  #1 \right\}  }

\global\long\def\Imag#1{\mathrm{Im}\left\{  #1 \right\}  }

\global\long\def\Real#1{\mathrm{Re}\left\{  #1 \right\}  }

\global\long\def\db{\!\not\!\! D\,}

\global\long\def\gesim{\,{\raise-3pt\hbox{$\sim$}}\!\!\!\!\!{\raise2pt\hbox{$>$}}\,}
\global\long\def\then{{\quad\Rightarrow\quad}}
\global\long\def\lcal{{\cal L}}
\global\long\def\mcal{{\cal M}}
\global\long\def\bBB{{\mathbbm B}}
\global\long\def\sigbf{\bm{\sigma}}
\global\long\def\gev{\hbox{GeV}}
\global\long\def\tev{\hbox{TeV}}
\global\long\def\vevof#1{\left\langle #1\right\rangle }
\global\long\def\up#1{^{\left( #1 \right) }}
\global\long\def\inv#1{\frac{1}{#1}}
\global\long\def\su#1{{SU(#1)}}
\global\long\def\ui{U(1)}

%
\begin{comment}
FTUV-09-1120, IFIC-09-61, UCRHEP-T478
\end{comment}
{}

\title{A model for right-handed neutrino magnetic moments}

\author[valencia]{Alberto Aparici }

\author[valencia]{Arcadi Santamaria }

\author[riverside]{José Wudka}

\address[valencia]{Departament de Física Teòrica, Universitat de València\\
 and IFIC, Universitat de València-CSIC\\
Dr. Moliner 50, E-46100 Burjassot (València), Spain}

\address[riverside]{Department of Physics and Astronomy, University of California, Riverside
CA 92521-0413, USA}
\begin{abstract}
A simple extension of the Standard Model providing Majorana magnetic
moments to right-handed neutrinos is presented. The model contains,
in addition to the Standard Model particles and right-handed neutrinos,
just a singly charged scalar and a vector-like charged fermion. The
phenomenology of the model is analysed and its implications in cosmology,
astrophysics and lepton flavour violating processes are extracted.
If light enough, the  charged particles responsible for the right-handed
neutrino magnetic moments could copiously be produced at the LHC. \end{abstract}
\maketitle

\section{Introduction}

In ref.~\cite{Aparici:2009fh} we studied the most general effective
Lagrangian built with the Standard Model (SM) fields plus right-handed
neutrinos up to operators of dimension five. We found this Lagrangian
contains only three nonrenormalizable operators, one of them being
the well known Weinberg operator~\cite{Weinberg:1979sa} which only
involves the SM lepton doublets and the Higgs doublet. The other two
contain an interaction of right-handed neutrinos with the SM Higgs
doublet and a Majorana electroweak moment for the right-handed neutrinos.
This last operator is particularly interesting and can have a variety
of phenomenological consequences in cosmology, astrophysics and at
colliders~\cite{Aparici:2009fh}. Of course, it is interesting to
have explicit models in which these nonrenormalizable interactions
arise naturally because one can use them to check the general features
of the effective Lagrangian approach and extend them outside the realm
of validity of the effective field theory. This is especially important
if the particles responsible for the new interactions are light enough
as to be produced at the  next generation of colliders.

Here we present a very simple model which gives rise to right-handed
neutrino electroweak moments; it includes, in addition to the SM fields
and the right-handed neutrinos, a charged scalar singlet and a charged
singlet vector-like fermion. We obtain the tree level and one-loop
contributions to the dimension five effective Lagrangian, and in particular
we compute the contribution to the right-handed neutrino electroweak
moments. We perform a thorough phenomenological analysis of the model,
paying special attention to the case in which the new charged particles
are light enough to be produced at the Large Hadron Collider (LHC).
Thus, in section~\ref{sec:The-model} we define the model and compute
the one-loop contribution to the electroweak moment of right-handed
neutrinos. The simplest version of the model, in which several couplings
are set to zero by using global symmetries, contains stable charged
massive particles (CHAMPs) which are strongly disfavoured from cosmological
and astrophysical considerations. To avoid such problems we extend
minimally the model by allowing a soft breaking of the symmetries,
which is enough to induce CHAMP decays; such decays are studied in
section~\ref{sub:broken-symmetry}. The model also induces some tree-level
lepton flavour violating (LFV) processes like $\mu\rightarrow3e$
which are studied in section~\ref{sub:LFV}. In section~\ref{sec:nur-higgs-operator}
we discuss briefly the one-loop contributions of the model to the
effective Higgs-$\nu_{R}$ operator. In section~\ref{sec:colliders}
we compute the production cross section of the  charged particles
at the LHC and discuss their observability as a function of their
masses. Finally, in section~\ref{sec:Conclusions} we present our
conclusions.

\section{The model\label{sec:The-model}}

As discussed in ref.~\cite{Aparici:2009fh} the most general dimension
five interactions among SM fields and three right-handed neutrinos
can be written as%
\footnote{The reader should note a difference in notation respect to \cite{Aparici:2009fh},
where we used $\nu^{\prime}$ to denote the neutrino flavor eigenfields.
As in the present work we are not going to discuss the diagonalisation
of the neutrino mass matrices we will just use $\nu$ to represent
the flavor eigenfields.%
} \begin{equation}
\mathcal{L}_{\mathrm{5}}=\overline{\nu_{R}^{c}}\zeta\sigma^{\mu\nu}\nu_{R}B_{\mu\nu}+\left(\overline{\tilde{\ell}}\phi\right)\chi\left(\left.\tilde{\phi}\right.^{\dagger}\ell\right)-\left(\phi^{\dagger}\phi\right)\overline{\nu_{R}^{c}}\xi\nu_{R}+\mathrm{h.c.}\label{eq:L5}\end{equation}
where $\ell={\nu_{L} \choose e_{L}}$ denotes the left-handed lepton
isodoublet, $e_{R}$ and $\nu_{R}$ the corresponding right-handed
isosinglets, and $\phi$ the scalar isodoublet (family and gauge indices
will be suppressed when no confusion can arise). The charge-conjugate
fields are defined as $e_{R}^{c}=C\bar{e}_{R}^{T},\:\nu_{R}^{c}=C\bar{\nu}_{R}^{T}$
and $\tilde{\ell}=\epsilon C\bar{\ell}^{\, T},\,\tilde{\phi}=\epsilon\phi^{*}$
where $\epsilon=i\sigma_{2}$ acts on the $SU(2)$ indices. The hypercharges
assignments are $\phi:\,1/2$, $\ell:-1/2$, $e_{R}:-1$, $\nu_{R}:\,0$.
The $SU(2)$ and $U(1)$ gauge fields are denoted by $W$ and $B$
respectively (gluon and quarks fields will not be needed in the situations
considered below). The couplings $\chi$, $\xi$, $\zeta$ have dimension
of inverse mass, which is associated with the scale of the heavy physics
responsible for the corresponding operator. $\chi$, and $\xi$ are
complex symmetric $3\times3$ matrices in flavour space, while $\zeta$
is a complex antisymmetric matrix proportional to the right-handed
neutrino electroweak moments.

The different terms in eq.~\eqref{eq:L5} and their phenomenological
consequences were discussed in~\cite{Aparici:2009fh}. %
\begin{comment}
The term involving $\chi$ was first described by Weinberg~\cite{Weinberg:1979sa}
and provides a Majorana mass for the left-handed neutrino fields plus
various lepton-number-violating neutrino-Higgs interactions; this
effective operator is the same one obtains when considering generic
see-saw models. The term involving $\zeta$ describes electroweak
moment couplings of the right-handed neutrinos discussed in ref.~\cite{Aparici:2009fh}.
The term involving $\xi$ gives a contribution to the right-handed
neutrino Majorana mass but also has the interesting feature that could
induce invisible Higgs decays~\cite{Aparici:2009fh}. $\chi$ and
$\xi$ appear naturally at tree level in many extensions of the SM. 
\end{comment}
{}Here we are more interested in models that could give rise to $\zeta$.
This can only occur at the one-loop level and the models should necessarily
involve either a scalar-fermion pair with opposite (non-zero) hypercharges
and having Yukawa couplings with both $\nu_{R}$ and $\nu_{R}^{c}$,
or a vector-fermion pair with the same properties. Here we will consider
only the first (simpler) possibility. Thus we enlarge the SM by adding
a negatively charged scalar singlet $\omega$, $Y(\omega)=-1$, and
one negatively charged vector-like fermion $E$ (two chiralities and
no generation indices) also with $Y(E)=-1$.

We can then write the Lagrangian as

\begin{equation}
\mathcal{L}=\mathcal{L}_{\mathrm{SM}}+\mathcal{L}_{\mathrm{NP}}\,\,,\label{eq:lagrangian}\end{equation}
where $\mathcal{L}_{\mathrm{SM}}$ is the SM Lagrangian while the
new physics Lagrangian, $\mathcal{L}_{\mathrm{NP}}$, collects all
the terms containing any of the new particles, including among them
the right-handed neutrinos. We write $\mathcal{L}_{\mathrm{SM}}$
as \begin{equation}
\mathcal{L}_{\mathrm{SM}}=i\overline{\ell}\db\ell+i\overline{e_{R}}\db e_{R}+(\overline{\ell}Y_{e}e_{R}\,\phi+\mathrm{h.c.})+\cdots\label{eq:SMlagrangian}\end{equation}
with $Y_{e}$ the Yukawa couplings of charged leptons which are completely
general $3\times3$ matrices in flavour space; the dots represent
SM gauge boson, Higgs boson and quark kinetic terms, quark Yukawa
interactions and the SM Higgs potential. We divide the  new physics
contribution, $\mathcal{L}_{NP}$, in different terms:\begin{equation}
\mathcal{L}_{NP}=\mathcal{L}_{K}+\mathcal{L}_{Y}-V_{NP}+\mathcal{L}_{Extra}\,\label{eq:LNP}\end{equation}
$\mathcal{L}_{K}$ describes the kinetic terms of the  new particles

\begin{equation}
\mathcal{L}_{K}=D_{\mu}\omega^{\dagger}D^{\mu}\omega+i\overline{E}\db E-m_{E}\bar{E}E+i\bar{\nu}_{R}\sla\partial\nu_{R}-\left(\half\overline{\nu_{R}^{c}}M_{R}\nu_{R}\,+\mathrm{h.c.}\right)\label{eq:lkinetic}\end{equation}
with $M_{R}$ the Majorana mass term of right-handed neutrinos, which
is a complex symmetric matrix in flavour space. $\mathcal{L}_{Y}$
contains the standard Yukawa interactions of right-handed neutrinos
and the Yukawa couplings of right-handed neutrinos with the particles
needed to generate the electroweak moments:

\begin{equation}
\mathcal{L}_{Y}=\overline{\ell}Y_{\nu}\nu_{R}\,\tilde{\phi}+\overline{\nu_{R}^{c}}h^{\prime}E\,\omega^{+}+\overline{\nu_{R}}hE\,\omega^{+}\,+\mathrm{h.c.}\label{eq:lyukawa}\end{equation}
$Y_{\nu}$ is a general $3\times3$ complex matrix and, if there is
just one $E$, $h$ and $h^{\prime}$ are vectors in generation space.
The $\omega$ contributions to the scalar potential are\begin{equation}
V_{NP}=m_{\omega}^{\prime2}|\omega|^{2}+\lambda_{\omega}|\omega|^{4}+2\lambda_{\omega\phi}|\omega|^{2}\phi^{\dagger}\phi\,,\qquad m_{\omega}^{2}=m_{\omega}^{\prime2}+\lambda_{\omega\phi}v^{2}\end{equation}
Where $v$ is the vacuum expectation value of the Higgs doublet, $\langle\phi^{\dagger}\phi\rangle=v^{2}/2$,
and the $\lambda$'s are quartic scalar couplings. We assume $\lambda,\lambda_{\omega}>0$
and $\lambda\lambda_{\omega}>\lambda_{\omega\phi}^{2}$ to insure
global (tree-level) stability, as well as $m_{\omega}^{2}>0$ in order
to preserve $U(1)_{\mathrm{em}}$. It is important to remark that
with only one Higgs doublet there cannot be trilinear couplings between
the doublet and the singlet, $\omega$. Then, the potential has two
independent $U(1)$ symmetries, one for the singlet and one for the
doublet.

In addition, the SM symmetries allow the following Yukawa couplings
and mass terms

\begin{equation}
\mathcal{L}_{Extra}=\bar{E}_{L}\kappa e_{R}+\overline{\ell}Y_{E}E_{R}\,\phi+\overline{\tilde{\ell}}f\ell\omega^{+}+\bar{e}_{R}f^{\prime}\nu_{R}^{c}\omega+\mathrm{h.c.}\label{eq:lextra}\end{equation}
 which can be set to zero by imposing a discrete symmetry which affects
only the new  particles \begin{equation}
E\rightarrow-E\,,\qquad\omega\rightarrow-\omega\end{equation}
In this case \emph{all}  low-energy physics effects will be loop generated\cite{Wudka:2005yp}.
Notice that the resulting Lagrangian has a larger continuous symmetry
\begin{equation}
E\rightarrow e^{i\alpha}E\,,\qquad\omega\rightarrow e^{i\alpha}\omega\label{eq:global-symmetry}\end{equation}
 which is not anomalous, therefore there is a charge, carried only
by $E$ and $\omega$ which is exactly conserved. In that case, the
lightest of the $E$ or $\omega$ will be completely stable becoming
a CHAMP, which could create serious problems in standard cosmology
scenarios. However, such problems can easily be evaded by allowing
some of the terms in eq.~\eqref{eq:lextra}. We will return to this
issue after verifying that the model indeed generates a right-handed
neutrino magnetic moment.

\subsection{The $\nu_{R}$ magnetic moment\label{sub:nur-magmo}}

\begin{figure}
\begin{centering}
\includegraphics[width=1\columnwidth]{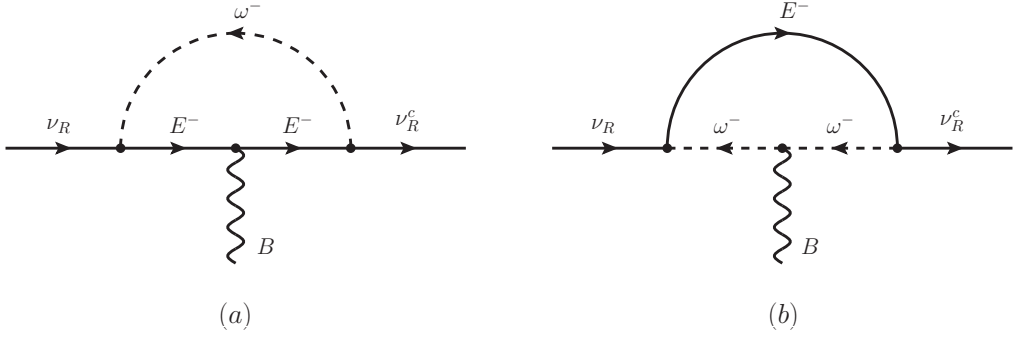}
\par\end{centering}

\caption{Contributing diagrams to the right-handed neutrino electroweak moment.\label{fig:Magmo-diagrams}}

\end{figure}

In the model considered we have two diagrams, depicted in figure~\ref{fig:Magmo-diagrams},
contributing to the $\nu_{R}$ Majorana electroweak moment: a) loop
with the $B$ gauge boson attached to the $E$ and b) loop with the
$B$ gauge boson attached to the scalar $\omega$. 

For $M_{R}\ll m_{E},m_{\omega}$ we can neglect all external momenta
and masses and the calculation of the diagrams simplifies considerably.
The final result can be cast as a contribution to the effective magnetic
moment operator in eq.~\eqref{eq:L5}. We find

\begin{equation}
\zeta_{ij}=\frac{g^{\prime}f(r)}{(4\pi)^{2}4m_{E}}\left(h_{i}^{\prime}h_{j}^{*}-h_{j}^{\prime}h_{i}^{*}\right)\end{equation}
with $r=(m_{\omega}/m_{E})^{2}$, $g^{\prime}$ the $B_{\mu}$ gauge
coupling and\begin{equation}
f(r)=\frac{1}{1-r}+\frac{r}{(1-r)^{2}}\log(r)\rightarrow\left\{ \begin{array}{c}
1\,,\,\,\,\,\,\,\,\,\,\,\: r\ll1\\
1/2\,,\,\,\,\,\,\,\,\,\,\,\: r=1\\
(\log(r)-1)/r\,,\qquad r\gg1\end{array}\right.\label{eq:def-fr}\end{equation}

%
\begin{comment}
Thus we have \begin{equation}
\zeta_{ij}=\frac{g^{\prime}}{(4\pi)^{2}4m_{E}}\left(h_{i}^{\prime}h_{j}^{*}-h_{j}^{\prime}h_{i}^{*}\right)\,,\qquad m_{E}\gg m_{\omega},\end{equation}

\begin{equation}
\zeta_{ij}=\frac{g^{\prime}}{(4\pi)^{2}8m_{E}}\left(h_{i}^{\prime}h_{j}^{*}-h_{j}^{\prime}h_{i}^{*}\right)\,,\qquad m_{E}=m_{\omega},\end{equation}

\begin{equation}
\zeta_{ij}=\frac{g^{\prime}m_{E}}{(4\pi)^{2}4m_{\omega}^{2}}\left(\log\frac{m_{\omega}^{2}}{m_{E}^{2}}-1\right)\left(h_{i}^{\prime}h_{j}^{*}-h_{j}^{\prime}h_{i}^{*}\right)\,,\qquad m_{E}\ll m_{\omega}\end{equation}

\end{comment}
{}For an estimate we can take, for instance, $m_{\omega}=m_{E}$, and
$\left(h_{i}^{\prime}h_{j}^{*}-h_{j}^{\prime}h_{i}^{*}\right)=0.5$
while $g^{\prime}=\sqrt{\alpha4\pi}/c_{W}\approx0.35$, then $\zeta\approx10^{-4}/m_{E}$
(for $m_{E}\gg m_{\omega}$ there will be a factor $2$ enhancement
and for $m_{E}\ll m_{\omega}$ there will be a suppression by roughly
a factor $(m_{E}/m_{\omega})^{2}$); these values are in agreement
with the estimates obtained using effective field theory. In terms
of $\Lambda_{NP}\equiv1/\zeta\,$ we have $\Lambda_{NP}=10^{4}m_{E}$.
Present bounds from LEP and Tevatron give $m_{E}\gtrsim100\,\mathrm{GeV}$,
which imply $\Lambda_{NP}\gtrsim10^{6}\,\mathrm{GeV}$. This can be
compared with direct bounds that can be set on the right-handed neutrino
electroweak moments derived in~\cite{Aparici:2009fh} . As expected,
collider limits on $E$ production are much more restrictive than
collider limits derived from the induced electroweak moment interaction.
After all, the electroweak moment interaction is generated at one
loop. However, if the right-handed neutrinos are relatively light
(below $10\,\mathrm{MeV}$) bounds from transition magnetic moments
coming from supernova cooling (which are $\Lambda_{NP}\gtrsim4\times10^{6}\,\mathrm{GeV}$)
or red giant cooling (which are $\Lambda_{NP}\gtrsim4\times10^{9}\,\mathrm{GeV}$
for $m_{N}\lesssim10\,\mathrm{keV}$) can be much stronger.

\subsection{E or $\omega$ as CHAMPs\label{sub:champs}}

The model as described so far contains only the couplings necessary
to generate the right-handed neutrino Majorana electroweak moments.
But it is clear that the trilinear vertices $\bar{\nu}_{R}E\omega^{\dagger}$
and $\bar{\nu}_{R}^{c}E\omega^{\dagger}$ alone cannot induce decays
for both the $E$ and the $\omega$. The lightest of the two will
remain stable and could then accumulate in the galaxy clusters, appearing
as electrically charged dark matter. The idea that dark matter could
be composed mostly of charged massive particles was proposed in~\cite{DeRujula:1989fe,Dimopoulos:1989hk}
and it is strongly constrained from very different arguments\cite{Basdevant:1989fh,Hemmick:1989ns,Yamagata:1993jq,Gould:1989gw,Chivukula:1989cc}.
One might still consider the possibility of having massive stable
$E$ or $\omega$ particles within the reach of the LHC, but with
a cosmic abundance lower than the one required for dark matter. Unfortunately,
such scenario seems also to be excluded: if one assumes, as in \cite{DeRujula:1989fe},
that the $E$'s and $\omega$'s were produced in the early universe
through the standard freeze-out mechanism \cite{Wolfram:1978gp},
the bounds from interstellar calorimetry \cite{Chivukula:1989cc}
and terrestrial searches for super-heavy nuclei \cite{Hemmick:1989ns,Yamagata:1993jq}
completely close the window of under-TeV CHAMP abundances.

There is, however, a way to escape all these bounds. A  recent paper
\cite{Chuzhoy:2008zy} notes that CHAMPs, if very massive or carrying
very small charges, are expelled from the galactic disk by the magnetic
fields. That situation prevents any terrestrial or galactic detection
and leaves room for CHAMPs to exist. The bound specifically states
that particles with $100(Q/e)^{2}\;\mathrm{TeV}\lesssim m\lesssim10^{8}(Q/e)\;\mathrm{TeV}$
are depleted from the disk, and in fact our model (if we forbid the
terms in eq.~\eqref{eq:lextra}) does not fix the hypercharge of
$E$ and $\omega$, so they can be millicharged. Unfortunately, this
situation is not interesting for our purposes, for this kind of CHAMPs
would give rise to very small neutrino magnetic moments and wouldn't
show up in the future accelerators, either due to their heavy masses
or to their small couplings.

In conclusion, we need an additional mechanism for $E$ or $\omega$
decays. The easiest way to accomplish this is by allowing one or more
of the couplings in eq.~\eqref{eq:lextra}, which can be taken small,
if needed, by arguing that \eqref{eq:global-symmetry} is an almost
exact symmetry. We discuss one of the possibilities in section~\ref{sub:broken-symmetry}.
The scenario of decaying CHAMPs has, on its own, a number of advantages
and drawbacks. Some recent papers \cite{Jedamzik:2007qk,Jedamzik:2007cp}
have pointed out that the presence of a massive, charged and colourless
particle during the process of primordial nucleosynthesis might lead
to an explanation for the cosmic lithium problem. Also, the decay
of massive particles during nucleosynthesis could have a dramatic
influence in the final abundances of primordial elements, which provides
us with  bounds on the lifetime and abundance of CHAMPs that could
be useful.

\subsection{Allowing for CHAMP decays\label{sub:broken-symmetry}}

If the  particles have to decay the global symmetry \eqref{eq:global-symmetry}
has to be broken, and for that it is enough to allow some of the terms
in eq.~\eqref{eq:lextra}. For the sake of simplicity, we will consider
only the case where the symmetry is softly broken by $E_{L}$--$e_{R}$
mixing%
\footnote{Since this choice breaks \eqref{eq:global-symmetry} softly, none
of the other terms in eq.~\eqref{eq:lextra} need be introduced for
the model to remain renormalizable. %
}

\begin{equation}
\mathcal{L}_{\kappa}=\bar{E}_{L}\kappa e_{R}+\mathrm{h.c.}\label{eq:soft}\end{equation}

This term will induce decays of $E$ into SM particles much like the
heavy neutrino decays in seesaw models, since only this mixing links
the $E$ to the SM degrees of freedom. After diagonalisation of the
charged lepton mass matrix one obtains interactions that connect the
$E$ to $W+\nu$, $Z+\ell^{\pm}$ and $H+\ell^{\pm}$. As the current
bound on heavy charged leptons require that $m_{E}>100\,\mathrm{GeV}$,
the $W$ and $Z$ will be produced on-shell; the Higgs channel may
or may not be open depending on the actual value of the Higgs and
$E$ masses%
\footnote{Note that, as $U(1)_{\mathrm{em}}$ is not broken, flavour-changing
vertices involving a photon cannot appear at tree level; $\Gamma(E\rightarrow e\gamma)$
must be at least a one-loop effect, and thereby suppressed.%
}.

The $\omega$, on the other hand, has to decay through the Yukawa
$\bar{E}\nu_{R}\omega$ vertices; either directly to $E+\nu_{R}$
if $m_{\omega}>m_{E}$ or to $e+\nu_{R}$ suppressed by the mixing
$\kappa$. The simplest situation then arises if $m_{\omega}>m_{E}$,
for in that case the $\omega$'s will decay into on-shell $E$'s,
which in turn will decay in the aforementioned way. In what remains,
for simplicity, we shall restrict ourselves to this specific case.

\begin{figure}[h]
\begin{centering}
\includegraphics[width=0.8\columnwidth]{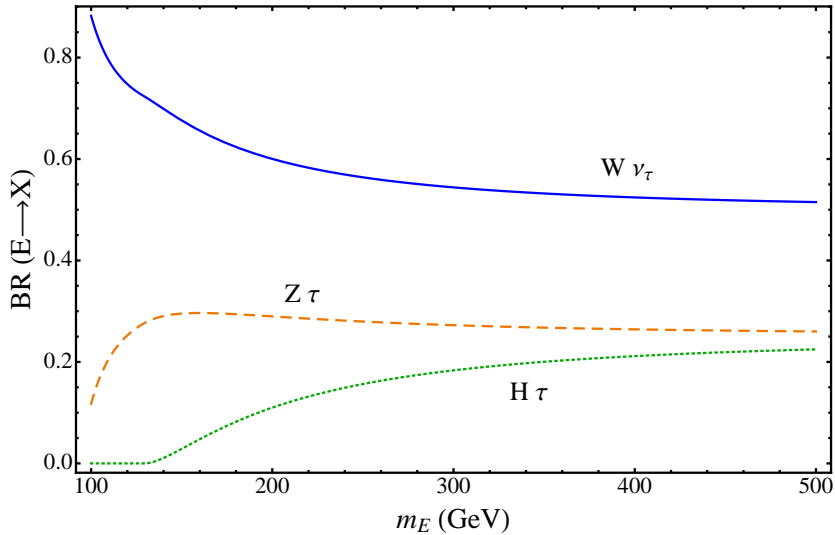}
\par\end{centering}

\caption{Dominant decay branching ratios of the vector-like fermion $E$. The
decays are suppressed by the mass of the charged leptons, thus we
have only represented decays into the third family. The Higgs boson
mass has been taken to the present best fit, $m_{H}=129\,\mathrm{GeV}$.\label{fig:BR-E}}

\end{figure}

In figure~\ref{fig:BR-E} we present the branching ratios for the
decays of the $E$. As the decays are controlled by the would-be Goldstone
part of the $W$ and $Z$ (and the Higgs boson if allowed kinematically)
they are always proportional to the Yukawa couplings of the charged
leptons; therefore, if all the $\kappa$'s are of the same order,
the $E$ will decay mainly to the leptons of the third family.  We
can see that for relatively low masses the dominant channel is $E\rightarrow W\nu_{\tau}$
while for very large masses the ratios tend to the equivalent-Goldstone
approximation: $0.5$ for the $W$ channel and $0.25$ for the $Z$
and $H$ channels. 

\begin{figure}[h]
\begin{centering}
\includegraphics[width=0.8\columnwidth]{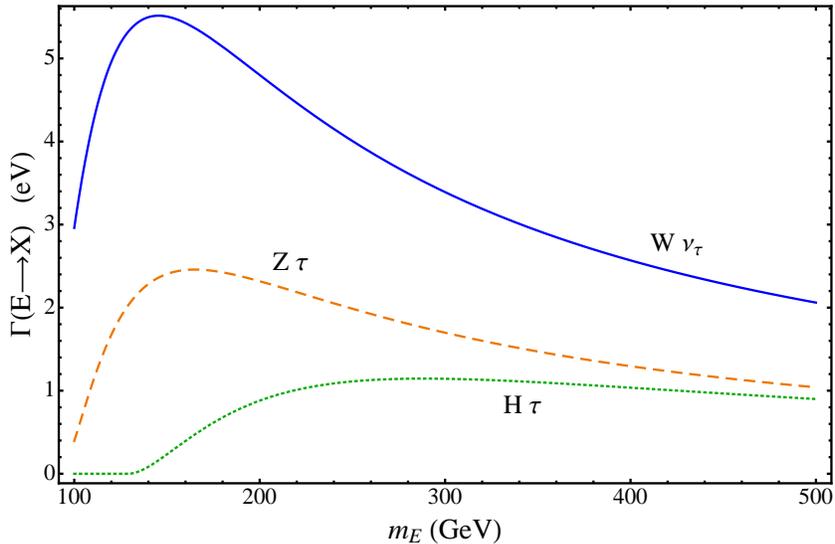}
\par\end{centering}

\caption{Dominant decay rates of the vector-like fermion $E$ with the same
assumptions made in figure~\ref{fig:BR-E}. For these estimates we
have taken $\kappa_{\tau}=1\,\mathrm{GeV}$.\label{fig:Gamma-E}}

\end{figure}

The decay rates of the $E$ fermion are presented in figure~\ref{fig:Gamma-E}
for $\kappa_{\tau}=1\,\mathrm{GeV}$. Notice that the rates decrease
for large $m_{E}$. This is because the decays proceed through the
mixing $E$--$\tau$ and this is suppressed by factors $m_{\tau}/m_{E}$;
thus the increase in phase space for large $m_{E}$ is compensated
by these factors. For the chosen value of $\kappa_{\tau}$ the decay
widths are of the order of the eV. For widths of this order of magnitude
the $E$'s will not be present at the time of primordial nucleosynthesis
and will not affect it. Note, however, that the decay rates depend
on $\kappa_{\tau}^{2}$, and $\kappa_{\tau}$ is relatively free,
thus the decay rates can vary in several orders of magnitude depending
on the value of $\kappa_{\tau}$. For $\kappa_{\tau}<10^{-7}\,\mathrm{GeV}$
the CHAMPs will affect nucleosynthesis and, as commented above, might
help to solve the cosmic lithium problem \cite{Jedamzik:2007qk,Jedamzik:2007cp}.
We also require $\kappa_{\tau}>10^{-16}\,\mathrm{GeV}$ to avoid CHAMPs
at the present epoch.

\subsection{Lepton Flavour Violating processes\label{sub:LFV}}

For general $\kappa$'s and Yukawa couplings $Y_{e}$, family lepton
flavour is not conserved; one might then worry about possible bounds
set by processes like $\mu\rightarrow3e$, $\mu\rightarrow e\gamma$
or $\tau\rightarrow3\mu$. We now determine whether the bounds on
those rare processes can impose restrictions on the parameters of
our model.

The easiest way to calculate the amplitudes for these processes is
by using an effective Lagrangian obtained by integration of the $E$
field. This integration is performed by using the equations of motion
for $E$ and expanding in powers of $1/m_{E}$ (for a detailed example
of the integration of a singly charged scalar see~\cite{Bilenky:1993bt}).
One then obtains 

\begin{equation}
\mathcal{L}\mathrm{_{LFV}}=-\frac{1}{m_{E}^{4}}\overline{e_{R}}\kappa\kappa^{\dagger}i\db^{3}e_{R}+\cdots\end{equation}
which, after the use of the equations of motion and spontaneous symmetry
breaking leads to a lepton flavour violating interaction of the $Z$
gauge boson with left-handed charged leptons, \begin{equation}
\mathcal{L}\mathrm{_{LFV}}=\frac{e}{2s_{W}c_{W}}Z_{\mu}\overline{e_{L}}C_{\mathrm{LFV}}\gamma^{\mu}e_{L}\;,\qquad C_{\mathrm{LFV}}\approx\frac{v^{2}}{2m_{E}^{4}}Y_{e}\kappa\kappa^{\dagger}Y_{e}^{\dagger}\,.\label{eq:LFV-lagragian}\end{equation}
$C_{\mathrm{LFV}}$ is a matrix in flavor space which is not, in general,
diagonal; therefore, eq.~\eqref{eq:LFV-lagragian} will induce processes
such as $\mu\rightarrow3e$ and $\tau\rightarrow3\mu$. Without loss
of generality we can take $Y_{e}$ diagonal with elements proportional
to the charged lepton masses; then we can estimate the branching ratio
for the $\mu\rightarrow3e$ process as

\begin{equation}
BR(\mu\rightarrow3e)=\frac{\Gamma(\mu\rightarrow3e)}{\Gamma(\mu\rightarrow e\nu\bar{\nu})}\approx\frac{\left|m_{e}\left(\kappa\kappa^{\dagger}\right)_{e\mu}m_{\mu}\right|^{2}}{m_{E}^{8}}\end{equation}
Our effective Lagrangian is an expansion in powers of $1/m_{E}$ which
could be compensated, in part, by $\kappa\kappa^{\dagger}$ factors
in the numerator; thus, for consistency, we should require $\kappa<m_{E}$
which allows us to establish an upper bound for the branching ratio.
Recalling also that the present limit on the mass of charged heavy
leptons is around $100\,\mathrm{GeV}$, and therefore we should have
$m_{E}>100\,\mathrm{GeV}$, we obtain\begin{equation}
BR(\mu\rightarrow3e)<\left(\frac{m_{\mu}m_{e}}{(100\,\mathrm{GeV)^{2}}}\right)^{2}<10^{-16}\end{equation}
to be compared with present bounds%
\footnote{All experimental limits are taken from\cite{Amsler:2008zzb}. %
} which are of the order of $10^{-12}$. If we apply the same reasoning
to $\tau\rightarrow3\mu$ we see that the branching ratio is enhanced
by a $(m_{\tau}/m_{e})^{2}$ factor\begin{equation}
R(\tau\rightarrow3\mu)\equiv\frac{\Gamma(\tau\rightarrow3\mu)}{\Gamma(\tau\rightarrow\mu\nu\bar{\nu})}<\left(\frac{m_{\tau}m_{\mu}}{(100\,\mathrm{GeV)^{2}}}\right)^{2}<10^{-10}\end{equation}
which is still under the present sensitivity for this ratio, which
is about $10^{-7}$. 

Another very restrictive process is $\mu\rightarrow e\gamma$, which
is bounded at the $10^{-11}$ level, $BR(\mu\rightarrow e\gamma)<1.2\times10^{-11}$.
This limit will be improved in a close future by the MEG experiment
by two orders of magnitude~\cite{Ritt:2006cg}. However, this process
can only arise at one loop and it is suppressed by loop factors; therefore,
we do not expect stringent bounds from it. The contributions to the
oblique parameters are suppressed by powers of the fermions masses
and are too small to be observed at the currently available precision.

Finally, $\mu$--$e$ conversion in nuclei also provides strong limits
in general; for instance, $\mu$--$e$ conversion on $\mathrm{Ti}$
gives $\sigma(\mu^{-}\mathrm{Ti}\rightarrow e^{-}\mathrm{Ti})/\sigma(\mu^{-}\mathrm{Ti}\rightarrow\mathrm{capture})<4.3\times10^{-12}$.
In our model, the process is induced by exactly the same interaction
\eqref{eq:LFV-lagragian} that gives $\mu\rightarrow3e$, and we again
do not expect, at present, a strong bound from $\mu$--$e$ conversion.
However, given the future plans to improve the limits by several orders
of magnitude, then perhaps $\mu$--$e$ conversion will provide the
best bound for LFV processes in this model. In any case, current data
on LFV processes cannot constrain this mechanism for $E$ decays.

\section{The $\nu_{R}$ mass and the effective Higgs boson interaction with
$\nu_{R}$\label{sec:nur-higgs-operator}}

The model we have discussed contains several sources of lepton number
non-conservation: the right-handed neutrino Majorana mass and the
$h$ and $h^{\prime}$ couplings (if both of them are different from
zero). Then it is interesting to ask what is the natural size of the
right-handed neutrino Majorana masses, since, even if they are set
to zero by hand, radiative corrections involving couplings that do
not conserve lepton number will generate them. In fact, by removing
the photon line in the diagrams that give rise to the electroweak
moments, figure~\ref{fig:Magmo-diagrams}, one obtains a renormalization
of the right-handed neutrino Majorana mass. The diagrams are logarithmically
divergent and give corrections of the type

\begin{equation}
\delta M_{R}\sim\frac{h^{\prime}h}{(4\pi)^{2}}m_{E}\,\end{equation}
(if the scalar $\omega$ is much heavier than the $E$, this contribution
will have an extra suppression $(m_{E}/m_{\omega})^{2}$). It is then
natural to require $M_{R}\gtrsim h^{\prime}hm_{E}/(4\pi)^{2}$. Of
course these type of contributions can be renormalized into $M_{R}$
which, after all, is a free parameter of the theory. 

\begin{figure}
\begin{centering}
\includegraphics[width=0.5\columnwidth]{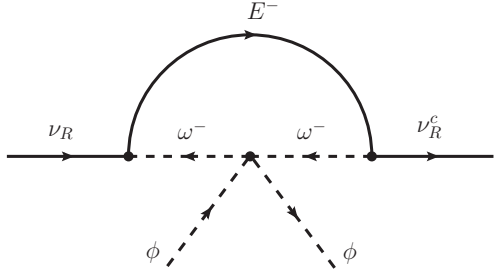}
\par\end{centering}

\caption{Diagram contributing to the $\left(\phi^{\dagger}\phi\right)\overline{\nu_{R}^{c}}\xi\nu_{R}$
operator.\label{fig:Higgs-nur}}

\end{figure}

In addition, similar diagrams with a vertex $(\phi^{\dagger}\phi)|\omega|^{2}$
attached to the $\omega$ field (see figure~\ref{fig:Higgs-nur})
give a finite contribution to the $\left(\phi^{\dagger}\phi\right)\overline{\nu_{R}^{c}}\xi\nu_{R}$
operator that cannot be avoided. A simple calculation gives

\begin{equation}
\xi_{ij}=\frac{\lambda_{\omega\phi}f_{\phi}(r)}{(4\pi)^{2}4m_{E}}\left(h_{i}^{\prime}h_{j}^{*}+h_{j}^{\prime}h_{i}^{*}\right)\end{equation}
where $f_{\phi}(r)$ can be written in terms of $f(r)$, defined in
eq.~\eqref{eq:def-fr}: $f_{\phi}(r)=4f(1/r)/r$. After spontaneous
symmetry breaking this operator gives additional contributions to
the right-handed Majorana neutrino mass\begin{equation}
\delta M_{R}\sim\frac{\lambda_{\omega\phi}h^{\prime}hv^{2}}{(4\pi)^{2}4m_{E}}\end{equation}
Therefore, at least, one should require \begin{equation}
M_{R}>\frac{\lambda_{\omega\phi}h^{\prime}hv^{2}}{(4\pi)^{2}4m_{E}}\sim\frac{\lambda_{\omega\phi}h^{\prime}h}{(4\pi)^{2}}100\,\mathrm{GeV\sim1\,}\mathrm{MeV}\end{equation}
where we took $h^{\prime}=h=\lambda_{\omega\phi}=0.1$. By taking
smaller couplings, smaller right-handed neutrino masses would be natural
(for instance for $h^{\prime}=h=\lambda_{\omega\phi}=0.01$ one obtains
$M_{R}>1\,\mathrm{keV}$).

\section{The Model at colliders\label{sec:colliders}}

\begin{figure}[h]
\begin{centering}
\includegraphics[width=0.8\columnwidth]{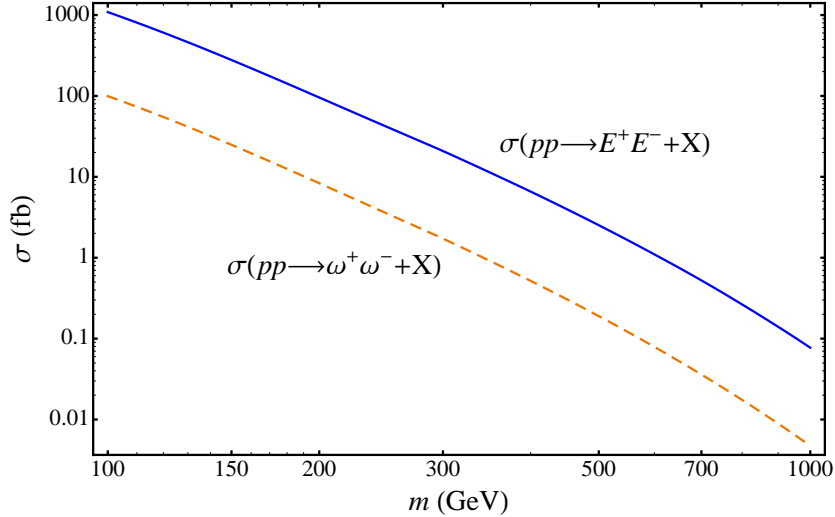}
\par\end{centering}

\caption{Production cross sections of the  charged particles at the LHC ($\sqrt{s}=14\,\mathrm{TeV})$
as a function of their masses. $m$ represents either $m_{E}$ or
$m_{\omega}$ depending on the process and $X$ represents that other
hadronic or leptonic products are expected in a proton-proton collision.
\label{fig:CS-LHC}}

\end{figure}

In spite of the fact that the new particles are $SU(2)$ singlets
and only have Yukawa couplings to right-handed neutrinos, they are
charged and can be copiously produced at the LHC, if light enough
($<1\,\mathrm{TeV}$), through the Drell-Yan process. 

The cross sections for proton-proton collisions can be computed in
terms of the partonic cross sections using the parton distribution
functions of the proton (for a very clear review see for instance~\cite{Campbell:2006wx});
in figure~\ref{fig:CS-LHC} we present the results%
\footnote{We have used the CTEQ6M parton distribution sets~\cite{Pumplin:2002vw}.
One could also include next-to-leading-order corrections by multiplying
by a $K$-factor which typically would change cross sections by $10-20\%$.
Results have been checked against the CompHEP program~\cite{Boos:2004kh,Pukhov:1999gg}.%
} for the production total cross sections at the LHC ($\sqrt{s}=14\,\mathrm{TeV})$
as a function of the $E$ and $\omega$ masses, $m_{E}$ and $m_{\omega}$
(both represented by $m$ in the figure). Since the  particles are
produced by $\gamma$ and $Z$ exchange, there are no unknown free
parameters except the masses of the particles. We see that cross sections
from $1\:\mathrm{fb}$ to $1\,\mathrm{pb}$ are easily obtained for
the production of $E$ for masses between $700\,\mathrm{GeV}$ and
$100\,\mathrm{GeV}$. For the same masses the production cross section
for $\omega$ is roughly one order of magnitude smaller.

Once produced in pairs, the  particles have to be detected and identified.
The characteristic signatures for this identification are very different
depending on the lifetimes of the  particles, mostly because if the
$E$ and $\omega$ are long-lived they can be tracked directly in
the detectors or, at least, be identified through a displaced decay
vertex. The parameter relevant for this behavior is $\kappa$, the
$E-e$ mixing.

For $\kappa\lesssim1\,\mathrm{MeV}$, the $E$'s will have decay lengths
roughly over $1$~centimeter%
\footnote{Note that there's room in the parameter space for this kind of effects
even if one requires that CHAMPs do not affect the primordial nucleosynthesis,
for if $\kappa>100\,\mathrm{eV}$ all the $E$'s will have decayed
before nucleosynthesis.%
}, in fact, for $\kappa<0.2\,\mathrm{MeV}$, they will go through the
detector and behave as a heavy ionizing particle. A lot of work has
been carried to analyse the signatures of CHAMPs inside the detector
(see, for example, \cite{Fairbairn:2006gg}, and \cite{Chen:2009gu}
for a recent improvement), and also displaced vertices have been discussed
(see, for example, \cite{Franceschini:2008pz,deCampos:2008re}).
If $\kappa>1\,\mathrm{MeV}$ the $E$'s will decay near the collision
point and behave as a fourth generation charged lepton. 

Discovering the $\omega$'s can be much harder, because they will
be produced at a significantly lower rate and the signatures of their
decays depend strongly on the details of the model. In the $m_{\omega}>m_{E}$
scenario, they will decay quickly into an $E$ and a heavy neutrino
(at least if we want $h$ and $h^{\prime}$ large enough to have significant
electroweak moments) and then one has to rely again on the detection
of $E$'s unless the heavy neutrino provides a cleaner signal, which
is unlikely. In any case, we think that the $E$'s, produced in a
much greater number, should be considered the signature of this model,
and perhaps the doorway to understand the $\omega$ and heavy neutrino
decays.

\section{Conclusions\label{sec:Conclusions}}

We have presented a simple model that generates right-handed neutrino
magnetic moments and studied its phenomenology. The simplest version
of the model contains CHAMPs (charged massive stable particles) which
could present some problems with standard cosmological scenarios.
These problems can easily be evaded by allowing additional couplings
in the Lagrangian. The model can then give rise to various LFV processes
at tree level such as $\mu\rightarrow3e$; however, we have verified
that the rates of these processes are strongly suppressed and are
well below present and near-future experimental constraints.

The same interactions that generate the right-handed neutrino magnetic
moments will also generate, at one loop, the last operator in eq.~\eqref{eq:L5}
which provides a  lepton number non-conserving interaction between
neutrinos and the SM Higgs boson. This interaction gives an additional
contribution to the right-handed neutrino Majorana mass; it is also
interesting because could lead to an invisible Higgs decay~\citep{Aparici:2009fh}.
We have computed it and discussed some of its consequences.

Finally, since the  particles responsible for the right-handed neutrino
magnetic moment are charged, if light enough they can copiously be
produced at the LHC through the Drell-Yan process. We found that the
cross sections for Drell-Yan production of $E$'s range from $1\:\mathrm{fb}$
to $1\,\mathrm{pb}$ for masses between $700\,\mathrm{GeV}$and $100\,\mathrm{GeV}$.
For the same range of masses the production cross section for $\omega$
is roughly one order of magnitude smaller. 

In short, we showed that a very simple model giving rise to right-handed
neutrino magnetic models compatible with all existing constraints
can easily be constructed. If the right-handed neutrinos are relatively
heavy ($\gtrsim10\,\mbox{MeV}$) bounds on $\nu_{R}$ magnetic moments
from red giants or supernovae do not apply~\cite{Aparici:2009fh}
and the  charged particles responsible for the magnetic moments could
be light enough as to be produced and detected at the LHC.

\section*{Acknowledgments}

This work has been supported in part by the Ministry of Science and
Innovation (MICINN) Spain, under the grant number FPA2008-03373, by
the {}``Generalitat Valenciana'' grant PROMETEO/2009/128, by the
European Union within the Marie Curie Research \& Training Networks,
MRTN-CT-2006-035482 (FLAVIAnet), and by the U.S. Department of Energy
grant No.~DE-FG03-94ER40837. A.A. is supported by the MICINN under
the FPU program.

%\bibliographystyle{h-elsevier}
%\bibliography{nur-model}

\end{document}